# Nanoscale membrane budding induced by CTxB on quasi-one component lipid bilayers detected by polarized localization microscopy

A. M. Kabbani and C. V. Kelly

## ABSTRACT

For endocytosis and exocytosis, membranes transition between planar, budding, and vesicular topographies through nanoscale reorganization of lipids, proteins, and carbohydrates. However, prior attempts to understand the initial stages of nanoscale bending have been limited by experimental resolution. Through the implementation of polarized localization microscopy (PLM), this manuscript reports the inherent membrane bending capability of cholera toxin subunit B (CTxB) in a quasi-one component supported lipid bilayers. Membrane buds were first detected with <50 nm radius, grew to >200 nm radius, and extended into longer tubules with dependence on the membrane tension and CTxB concentration. Compared to the concentration of the planar supported lipid bilayers, CTxB was >10x more concentrated on the positive curvature top and >25x more concentrated on the negative Gaussian curvature neck of the nanoscale membrane buds. These findings elucidate prior observations by correlating CTxB clustering and diffusion to CTxB-induced membrane bending. CTxB is frequently used as a marker for liquid-ordered lipid phases; however, the coupling between CTxB and membrane bending provides an alternate understanding of CTxB-induced membrane reorganization. Single-particle tracking was performed on both lipids and CTxB to reveal the correlation between single-molecule diffusion, CTxB accumulation, and membrane topography. Slowed lipid and CTxB diffusion was observed at the nanoscale buds locations, suggesting a local increase in membrane viscosity or molecular crowding upon membrane bending. These results suggest inherent CTxB-induced membrane bending as a mechanism for initiating CTxB internalization in cells that is initially independent of clathrin, caveolin, actin, and lipid phase separation.

**INTRODUCTION**

Membrane function is governed by the molecular organization, clustering, and interaction of its constituents. In particular, curvature-dependent reorganization has captured a growing interest as a mechanism for creating locally distinct membrane environments (1–3). In the presented study, we focus on the membrane bending effects of cholera toxin subunit B (CTxB) in a quasi-one component model membrane. Cholera toxin is a member of the AB5 toxin family that multivalently binds to GM1 and is most frequently used as the lipid raft marker in biophysical studies (4). CTxB-GM1 partitions with order-preferring lipids (5, 6), induces lipid phase segregation (6–8), and sorts to highly curvature regions (2, 3). GM1 plays a vital role in numerous biological functions including endocytosis (9), viral egress (10), disease propagation such as Alzheimer (11, 12), trafficking (13), and B cell signaling (14).

CTxB-GM1 assumes a sequence of macromolecular complexes between its initial membrane binding, clustering, and cellular internalization. Accordingly, numerous observations of multi-modal diffusion and nanoscale confinement of CTxB on living cells (15) and on synthetic bilayers (16, 17) have been reported. Even in the absence of coexisting lipid phases, CTxB exhibits multiple populations of diffusion rates and transient confinement in regions as small as 20 nm in radii (16, 17). On living cells, CTxB diffusion is independent of the diffusion of caveolin, clathrin, or glycosylphosphatidylinositol-linked proteins, which suggests the internalization of CTxB is initialized distinctly from conventional endocytotic processes (18–20).

Membrane inward vesiculation and tubulation has been observed in cells and synthetic vesicles upon exposure to Cholera toxin (10, 21). CTxB has been observed to sort to membranes of negative curvature for supported lipid bilayers (SLBs) on wavy glass (3), micron-scale nanoparticles (22), and membrane tethers (2). The capability of CTxB to bind to membranes in which both of the local principle curvatures are negative (*i.e.*, with a positive Gaussian curvature) is well established with CTxB-induced inward pits in giant unilamellar vesicles (GUVs) (23). This is supported by molecular dynamics simulations of the structurally similar Shiga toxin (23). However, the nanoscale details of CTxB intrinsically inducing membrane curvature, as necessary for endocytosis, and the capability of CTxB to bind to membranes with differing signs of principle curvatures remains uncertain.

We hypothesize that CTxB aggregates and internalizes as a result of its inherent physical effects on the membrane topography. Testing this hypothesis requires the use of an examination method that is able to resolve the colocalization of nanoscale membrane bending with CTxB. Polarized localization microscopy (PLM) combines single-molecule localization microscopy (SMLM) with polarized total internal reflection fluorescence microscopy (TIRFM) to detect nanoscale membrane orientation with super-resolution [cite PLM]. This technique distinguishes between membranes of varying orientation due to the differential excitation of membrane-confined fluorophores depending on the linear polarization of the incident excitation light. In particular, indocarbocyanine dyes (e.g., DiI) are photo-switchable probes (24) that maintain their fluorescence dipole moment in the plane of the membrane (25–27), such that membranes parallel to the coverslip are preferentially excited by incident s-polarized light, and membranes vertical to the coverslip are preferentially excited by incident p-polarized light. The robust identification of nanoscale membrane bending provided by PLM enables the correlation of membrane topography and molecular sorting on physiologically relevant length scales (<50 nm) with numerous technical advantages over other super-resolution techniques [cite PLM].

The microscopy setup for PLM permits simultaneous multicolor SMLM and single-particle tracking (SPT) of lipids and proteins. For example, Alexa Fluor dyes, conjugated to proteins, are flexible linkers that rotate sufficiently with negligible fluorescence excitation dependence on the illumination polarization. Such fluorescent dyes are common probes for imaging via direct stochastic optical reconstruction microscopy (dSTORM) (28). As demonstrated here, the multi-color, simultaneous combination of PLM and dSTORM enables the determination of membrane organization, molecular sorting, and single-molecule diffusion relative to membrane bending.

In this manuscript, we report the nanoscale organization and dynamics of CTxB relative to membrane bending events on a quasi-one component lipid bilayer with 99.4% 1-palmitoyl-2-oleoyl-sn-glycero-3-phosphocholine (POPC), 0.3% DiI, and 0.3% GM1. Using PLM, we found that the SLBs initially exhibit a flat uniform topology before the addition of CTxB, and the nanoscale membrane bending and bud formation occurred within 30 sec upon the addition of CTxB. Within that time, a subset of CTxB became clustered and the local diffusion of CTxB and DiI was slowed by (76 ± 4)% and (86 ± 10)% from the initial diffusion rate before CTxB addition, respectively. At later times after CTxB addition (>20 min), freely diffusing CTxB on planar bilayers, small accumulations of CTxB on nanoscale membrane buds, and rings of CTxB at larger membrane protrusions were simultaneously observed (Fig. 1). Single event analysis and spatially averaged correlation analysis demonstrated the strong interdependence of membrane structure, dynamics, and CTxB accumulation. In sum, these studies represent, to the best of our knowledge, the previously undetected phenomena of nanoscale membrane budding and tubulation by CTxB without the apparent need of lipid phase separation. PLM has enabled observing the effects of CTxB on spontaneous molecular sorting, immobilization, curvature, and tubule formation processes.

## MATERIALS and METHODS

### SLB formation

Giant unilamellar vesicles (GUVs) of primarily POPC (Avanti Polar Lipids, Inc.) with 0.3 mol% 1,1'-didodecyl-3,3,3',3'-tetramethylindocarbocyanine perchlorate (DiI, Life Technologies) and 0.3 mol% GM1 Ganglioside (Avanti Polar Lipids, Inc.) were prepared by electro-formation, as described previously[51]. Experiments were also repeated by using diphytanol phosphatidylcholine (DPhyPC, Avanti Polar Lipids, Inc.) instead of POPC and DiO or DiD instead of DiI with indistinguishable results. This composition yielded 110 nm$^2$ of bilayer per DiI or GM1 molecule. In brief, GUVs were formed by mixing lipids in chloroform and spreading them uniformly on a conducting indium tin oxide (ITO)-coated slide (Sigma-Aldrich) via spin coating. The resulting lipid films were dried under vacuum for one hour. A second ITO-coated slide and silicon spacer enclosed the dried lipids into an incubation chamber. A hydration buffer of 200 mM sucrose was added to the dried lipid films and the ITO slides were connected to either side of a sine wave function generator. The growth of the GUVs occurred over 3 hours at 55 °C with an alternating field at10 Hz and 2 V$_{rms}$. GUVs were stored at 55°C until use or discarded after 2 days. The interaction between the GUVs and plasma cleaned glass coverslips resulted in bursting of the GUVs and the formation of a continuous SLB over the glass.

### CTxB addition

CTxB was labeled with Alexa Fluor 647 or Alexa Fluor 488 prior to purchase from Thermo Fisher Scientific Inc. CTxB was added to the SLB for a final concentration of 1 μg/mL above the SLB to saturate all available GM1. After 0.5 min of incubation, the unbound CTxB was rinsed away. The time ($t$) is said to equal zero before CTxB was added, and otherwise, $t$ reports the time since the unbound CTxB was rinsed away. CTxB-Alexa Fluor 647 was used for all data shown below and indistinguishable results were obtained with CTxB-Alexa Fluor 488.

**Engineered membrane curvature**

Only in the indicated select experiments, membrane curvature was engineered prior to the addition of CTxB, as done previously [cite PLM]. In brief, 70 nm radius polystyrene nanoparticles of $\lambda_{ex}$ =488 nm (Fluoro-Max, Fisher Scientific) were exposed to a plasma cleaned coverslip of a glass bottom dish for 10 min to achieve a density of 0.02 NPs/µm$^2$. Glass bottom dishes were placed on a 55 °C hot plate for 5 min to ensure their stability on the coverslips. GUVs were fused to the coverslip and draped over the nanoparticles to create the engineered membrane curvature for greater consistency in membrane bud size in the indicated experiments. The index of refraction of polystyrene is 1.59 and may have resulted in a nanoscale shifting of the localization of single fluorophores, as discussed below.

**Imaging optics**

PLM was performed with an inverted IX83 microscope with Zero-Drift Correction and a 100x, 1.49NA objective (Olympus Corp.) on a vibration-isolated optical table. The high-NA objective permitted through-objective TIRFM. We have incorporated four continuous wave diode lasers at wavelengths 405, 488, 561, and 647 nm with at least 120 mW max power each for fluorescence excitation. The excitation polarization was rotated with a computer-controlled liquid crystal waveplate (Thorlabs Inc, LCC1111-A). Image acquisition was performed with an iXon-897 Ultra EMCCD camera (Andor Technology) proceeded by an OptoSplit IILS (Cairn Research) with emission filters (BrightLine, Semrock, Inc.), a 4-band notch filter (ZET405/488/561/640m, Chroma Corp.), and a 2x magnification lens. This setup provided high power (>80 mW) of linearly polarized fluorescence excitation and integrated computer control of all equipment via custom LabVIEW routines (National Instruments Corp.).

**Imaging procedure**

The sample was exposed to >80 mW of excitation light with $\lambda_{ex}$ = 561 (DiI) and $\lambda_{ex}$ = 647 nm (CTxB-AF647) simultaneously. Exposing the sample to high lasers powers for 3 s resulted in converting most of the fluorophores from their fluorescent state '*on*' to the transient non-fluorescent, dark state *'off'* to provide a steady state of well-separated fluorophore blinking. The '*on*' fluorophores were imaged at a density of less than one '*on*' fluorophore per 1 $\mu m^2$ per frame. Sequential movies were acquired with p-polarized total internal reflection (TIR) excitation at $\lambda_{ex}$ = 561 nm for pPLM and epifluorescence excitation at $\lambda_{ex}$ = 647 nm for dSTORM. 10,000 to 30,000 frames were acquired for each time point at a frame rate of 50 Hz on a region of interest with 18 ms acquisition per frame. Since CTxB was not illuminated with evanescent field of TIR, there was an increased probability of localizing the protein above the imaging focal plane compared to DiI that was illuminated via TIR. In this way, CTxB that were on the top of a membrane bud would be more likely to be localized than a DiI molecule on the top of a membrane bud.

**Imaging buffer**

PLM was performed on samples present in an oxygen-scavenging buffer (150 mM NaCl, 50 mM TRIS, 0.5 mg/mL glucose oxidase, 20 mg/mL glucose, 40 µg/mL catalase, and 1% β-marceptoethanol (BME) at pH 8). Buffer proteins were purchased from Sigma-Aldrich and salts were purchased from Fisher Scientific. These conditions maintain a low free oxygen concentration in the buffer to minimize non-reversible fluorophore bleaching and encourage transient fluorophore blinking, as is necessary for SMLM.

**Single-molecule localization**

The analysis of the raw, diffraction-limited images included low-pass Gaussian filtering, multi-emitter fitting routines, median background subtraction, lateral stage drift correction, and the fitting of each isolated fluorophore images via the ImageJ plug-in ThunderSTORM (29). ThunderSTORM provided the single fluorophore positions, localization uncertainty, and photon per fluorophores for further analysis. A threshold value 100 photons per fluorophores was used to keep only the bright localizations for further analysis. The localizations from CTxB and DiI excitation were analyzed separately to reconstruct separate super-resolution images for each color channel.

**Channels overlay**

The separate channel images were overlapped via custom-made MATLAB routine. A transformation was applied to precisely overlay the two half images relying on the locations of the TetraSpeck in both channels. The overlap was then optimized by maximizing the cross-correlation coefficient between the two channels as performed by other routines (30).

**Bud identification and size evaluation**

The detection of buds in each color channel was performed via custom-made MATLAB program that applies a mask and detects regions with > 3x the density of the average flat surrounding background bilayer. Each bud was fitted with a 2D Gaussian function for center estimation. The size of each bud ($r_{bud}$) was set equal to the mean distance from the bud center of all extra localizations due to the bud. This was calculated by taking into consideration the background from flat SLB localizations of uniform density ($\sigma$), the distance of each localization from the bud center ($r_i$), and a threshold distance that was significantly greater than $r_{bud}$ ($R$). Typically, $R$ = 400 nm but the following calculation is independent of the particular $R$ chosen. The number of extra localizations due to the presence of the bud ($N_{bud}$) is equal to the total number of localizations ($N_{all}$) within $r_i < R$ subtracted from the number of localizations expected within $R$ if no bud was present ($N_{SLB}$); $N_{SLB} = \pi R^2 \sigma = N_{all} - N_{bud}$. The mean $r_i$ expected for the flat SLB within $R$ is $2R/3$. By analyzing all collected localizations within $R$ and subtracting the expected localizations from the flat SLB, $r_{bud}$ is calculated according to

$$r_{bud} = \frac{\sum r_i}{N_{bud}} - \frac{2\pi\sigma R^3}{3N_{bud}}$$   (Eq. 1)

**Single-particle tracking**

SPT was performed on buds observed in both membrane and CTxB channels. The sequential localizations of single fluorophores were analyzed to reveal the diffusion rate of individual molecules versus membrane topography. The individual fluorophore trajectories $z$-projected onto the imaging $xy$-plane were identified with custom MATLAB code. Each fluorophore localizations were linked as a trajectory if they were in sequential frames, within a separation distance of 500 nm, and there was no alternative localization for linking within 1 μm. The single-molecule step lengths ($v$) were grouped based on their distance from the bud center, and their normalized distribution was fit via non-linear least squares method to a 2D Maxwell-Boltzmann distribution (Eq. 2) as would be expected for 2D Brownian diffusion.

$$P(v) = \frac{v}{2 D_{fit} \Delta t} e^{\frac{-v^2}{4 D_{fit} \Delta t}}.$$     (Eq. 2)

The localization imprecision ($\sigma$ = 20 nm) increased the apparent step lengths. Accordingly, the reported diffusion coefficient ($D$) was calculated from the fit of Eq. 2 according to $D = D_{fit} - \sigma^2/2/\Delta t$. The projection of the lipid trajectories onto the imaging $xy$-plane yielded a decrease in their apparent step lengths depending on the membrane tilt ($\theta$). The effects of membrane tilt are discussed below and in our companion manuscript [cite PLM].

Whereas diffusion coefficients are typically extracted from a trajectory by fitting the mean squared displacement versus $\Delta t$ for $\Delta t > 0$, this routine was not appropriate here. Fitting a longer trajectory to a single diffusion coefficient would have blurred the effects of curvature because each single trajectory samples both curved and flat membranes. Even with single-step analysis, a single step over 20 ms with a $D = 0.4$ $\mu m^2$/s, as is expected for DiI, would result in the averaging of the membrane environment over the expected 180 nm step length. With greater experimental sampling densities, rates, and precision, a more sophisticated analysis routine would be warranted.

## RESULTS

### CTxB induces membrane budding in SLBs

The reconstructed time-lapse dSTORM and pPLM images of CTxB and DiI revealed the initial protein accumulation and membrane budding processes, respectively. Within the first minute of CTxB addition to the membrane, some CTxB exhibited confinement on the flat bilayer, as demonstrated by a detectable accumulation of CTxB localizations without a significant increase in the local density of DiI localizations from pPLM (Fig. 2). After 1 min, the clusters of CTxB became co-localized with higher densities of DiI localizations as detected with pPLM. A local increase in DiI localizations obtained by pPLM represents areas in which the membrane would be more perpendicular to the microscopy coverslip, as would be expected for membrane bud. Membrane buds formed at the locations that CTxB accumulated, demonstrating the capability of CTxB to initiate and induce nanoscale membrane bending. PLM of DiI revealed both a continued growth in the size of the buds and the formation of new buds with continued CTxB exposure.

PLM allowed earlier visualization of membrane bending initiated by CTxB than was detectable by epifluorescence microscopy. Epifluorescence microscopy revealed a laterally uniform brightness of the DiI and CTxB on the SLB for the first 20 min of CTxB incubation. The formation of CTxB accumulations and lateral DiI variations became evident with epifluorescence microscopy after 20 min in some regions of the sample. Our interpretation of the lateral variations across SLBs and variations between SLBs created by differing methods is discussed below.

Regions of clustered localizations, observed in both DiI and CTxB channels, with localization density > 3x that of the flat membrane were identified as membrane buds. Histograms of bud sizes versus time from the pPLM of DiI and dSTORM of CTxB show buds increasing in number and size over time (Fig. 3). Additionally, this analysis further shows that CTxB accumulations precede membrane bending since the CTxB accumulations are larger and more numerous than the DiI accumulations, as is qualitatively shown in Fig. 2. Within this analysis, there was no clear separation of the buds into distinct stages of formation or growth into larger structures (*i.e.,* tubules).

**Some membrane buds grew into tubules**

A wide distribution of bud sizes was observed with increasing bud sizes present after longer CTxB incubation times (Fig. 3). The smallest buds displayed an apparently uniform distribution of CTxB across the bud where the specific distribution of CTxB on the bud was limited by the resolution of dSTORM (*i.e.*, Fig. 1A, *black arrows*). Intermediate size buds were observed in which the DiI localizations suggest a hemispherical membrane shape greater than 100 nm radius and CTxB preferentially localized at the bud neck (Fig. 1A, *white arrow*). These large hemispherical buds showed a radially decreasing density of DiI localizations rather than a ring of DiI localizations due to the anisotropic emission of the DiI orientationally confined within the membrane(31, 32) [cite PLM]. The largest membrane bending events observed were membrane tubules with the membrane protruding away from the glass coverslip by >3 μm (Fig. 4). In these membrane tubules, the ring of DiI localizations and the ring of CTxB localizations were both apparent.

**Buds vanish upon CTxB depletion**

CTxB concentration was a key factor in this membrane budding and tubule formation. Typically, the concentration of CTxB on the SLB was determined by saturating the 0.3% GM1 within the membrane and the CTxB concentration was apparently constant within 2 min of CTxB addition. However, in select experiments, the glass coverslip surrounding the patch of SLB was prepared as to encourage CTxB binding directly to the glass. In the first minutes following CTxB addition to the flat SLB, nanoscale membrane clusters were detected, as seen in all other experiments. However, the CTxB concentration was not constant over long times in this experiment. As CTxB laterally diffused on the membrane, it eventually came into close proximity with the perimeter of the SLB and the surrounding glass. Only in this experiment was CTxB observed to stick and accumulate on the glass surface (Fig. 5). This binding of CTxB to the glass caused a 15x decrease in CTxB concentration from 0.029 to 0.0018 localizations/nm$^2$ on the SLB. In contrast, the rate of DiI localizations showed no significant change over time. Meanwhile, the localization density of CTxB on the glass increased to 0.014 localizations/nm$^2$ a factor of 8.5x the initial density of 0.0017 localizations/nm$^2$. As a result of the decreasing CTxB concentration on the SLB, the number and size of the nanoscale buds decreased as shown in both CTxB and DiI localizations (Fig. 5). The buds and tubules disappeared from the membrane and the SLB returned to its original flat topography upon decreasing CTxB concentration, demonstrating the reversibility of membrane budding.

As a further control to confirm that the budding and tubulation processes were induced by CTxB binding to the membrane, these experiments were repeated without addition of CTxB. POPC/GM1/DiI membranes were imaged under identical conditions for 24 hours and no observable membrane deformations were detected.

**Single-molecule mobility varies with budding**

SPT was performed on both the DiI and CTxB as a function of location within the membrane buds. Single-fluorophores that stayed 'on' for sequential frames in the raw data collection were individually localized and linked to reveal the single-molecule mobility. The diffusive rate of DiI and CTxB on the planar SLB were measured to be (0.39 ± 0.05) µm$^2$/s and (0.09 ± 0.02) µm$^2$/s, respectively, with correcting for localization uncertainty in this single-step length analysis, as described above. The molecular diffusion was analyzed as a function of distance from the center of the membrane bud at various times of CTxB incubation (Fig. 6). Both the diffusion of DiI and CTxB demonstrated slowed diffusion rates through the *xy*-plane by $(76 \pm 4)\%$ and $(86 \pm 10)\%$ at the center of the membrane bud relative to the surrounding planar SLB. While a significant component of this perceived slowing of the single-molecule diffusion could be attributed to the membrane tilt, localization uncertainty ($\sigma = 20$ nm) and frame rate (50 Hz) limit, these conditions affect to at most a 40% slowing on nanoscale buds, as discussed below and in our companion manuscript [cite PLM]. The diffusion coefficients of DiI and CTxB were independent of the presence of buds for distances greater than 200 nm from buds centers, with a slight variation in CTxB incubation time that is correlated with bud size.

**Budding occurs with varying lipid types**

All budding experiments were repeated with DPhyPC replacing POPC as the primary membrane component to confirm that the particular lipids used here were not dominating this budding observation. DPhyPC and POPC are both liquid crystalline at room temperature, yet with a highly different molecular structure of their acyl chains. Indistinguishable curvature induction by CTxB was observed on SLBs formed with 99.4% DPhyPC, 0.3% DiI, and 0.3% GM1 as with 99.4% POPC, 0.3% DiI and 0.3% GM1. Further, experiments were also performed with varying membrane labels to ensure that probable lipid degradation or traces of imperfections were not the cause for such an observation. Membranes of 99.4% POPC, 0.3% GM1 and 0.3% DiO, DiI, or DiD was created and later CTxB-AF647 or CTxB-AF488 was added to the bilayers. CTxB induced membrane budding in all cases of varying lipids, lipid dyes and labeled CTxB.

**Membrane curvature is generated in unsupported bilayers**

CTxB-induced budding was reproduced on unfused GUVs. CTxB was introduced to POPC/GM1/DiI GUVs placed on an agarose-coated coverslip to prevent their rupture. Within 2 min of addition, CTxB, the suspended membranes bent and formed inward tubules. Most commonly, small vesicular invaginations coated with CTxB were observed (Fig. S1), similar to features reported previously (10, 21). Bending away from the leaflet exposed to CTxB was observed when possible; however, bending in this analogous direction were not possible for SLBs due to the close proximity (~2 nm) of the bilayer to the glass. When unable to bend away from the CTxB, the membrane buds and tubules grew outward, toward the CTxB, demonstrating a preference for CTxB to bind to positively curved membranes more than planar membranes.

**Quantifying CTxB sorting**

The buds formed by CTxB addition to planar SLBs were of sizes that varied from each other and varied with time (Fig. 3). To measure the partition coefficient of CTxB versus membrane curvature, a relatively static membrane curvature was engineered and the local CTxB concentration was observed. Membrane buds were engineered by draping SLBs over nanoparticles of known sizes. The ability to measure the CTxB distribution on multiple engineered buds of roughly of similar, known size enabled averaging between buds for lower noise in the experimental data and the fitting of the CTxB distribution to a predicted model of the membrane topography (Fig. 7). The radial density of CTxB observed on 25 separate nanoparticle-created membrane buds provides the experimental data (Fig. 7A). The predicted CTxB distribution was *z*-projected onto the *xy*-plane from a simulated membrane topography, (Fig. 7B) this enabled fitting the partition coefficients for CTxB versus membrane curvature. Fitting the model to the experimental results required incorporating the single-fluorophore localization imprecision, nanoparticle-induced inaccuracy, and imprecision in identifying the center of the nanoparticle; in addition to the curvature-dependent CTxB concentration. The curvature-dependent CTxB sorting was simplified to include just three concentrations: on the SLB ($[CTxB]_{SLB}$), over the top of the nanoparticle of 70 nm radii of curvature ($[CTxB]_{Top}$), and on the membrane neck with one principle radius of curvature equal to 20 nm ($[CTxB]_{Neck}$). Many combinations of these fitting parameters yielded similar quality fit to the experimental data. The mean and standard deviation of adequate fits to the experimental data yielded $[CTxB]_{Top}/[CTxB]_{SLB} = (12 \pm 4)$ and $[CTxB]_{Neck}/[CTxB]_{SLB} = (26 \pm 11)$ (Fig. 7C).

**DISCUSSION**

PLM is a novel microscopy technique that enables imaging membrane dynamics, organization, and topography simultaneously [cite PLM]. Since PLM requires no modification to the fluorescence emission path, it is trivially coupled to other super-resolution techniques, such as STORM and PALM. In addition to providing sub-diffraction-limited spatial resolution of membrane orientation, a strength of PLM is its high sensitivity to membrane bending and its ability to detect nanoscale membrane buds within a large planar membrane. Membrane buds are detected by PLM with higher signal-to-noise than any other comparable diffraction-limited or super-resolution optical technique [cite PLM].

In this manuscript, PLM was used to reveal nanoscale membrane curvature induced by membrane-bound CTxB. PLM provided direct, super-resolution time-lapse imaging of buds initiation and growth prior to being detectable by diffraction-limited imaging techniques. Each step in the progression from (1) binding of CTxB to the GM1 in a planar, quasi-one component SLB, (2) clustering of CTxB in the planar membrane, (3) inducing nanoscale membrane buds, and (4) the formation nanoscale membrane tubules protruding away from the coverslip were each detected. PLM has enabled this detection of nanoscale bud formation and the inherent membrane bending capability of CTxB that has previously gone unnoticed.

Super-resolution images of CTxB-induced membrane buds reveal Gaussian-like distributions of CTxB in the imaging $xy$-plane. These small membrane buds of radius ($r_{bud}$) equal $50 \pm 9$ nm displayed CTxB apparently bound upon the whole curved membrane of the bud, although the distribution of CTxB on the small bud was limited by the resolution of dSTORM. As the size of the buds increased, CTxB became most concentrated at the neck of the bud and yielded a "ring-like" structure of CTxB localizations when $r_{bud} > 100$ nm (Figs. 1 and S2).

**PLM distinguishes between buds and tubules**

Reconstructed pPLM images were able to distinguish between membrane buds and membrane tubules by the distribution of the DiI localizations; a heterogeneous population of bud sizes was calculated at each time point (Fig. 3). The confinement of the DiI within the lipid bilayers prevents free tumbling of the fluorophore, which is critical for PLM, but it also yields an anisotropic emission from the DiI and a systematic shift of the single-fluorophore image that is dependent on the membrane orientation and height (33). In particular, when the membrane is tilted 45° relative to the coverslip and 100 nm out of focus, the single-fluorophore image can be shifted by up to 50 nm. On membrane buds, the anisotropic emission effects on single DiI images systematically shifts the localizations towards the center of the bud and reduces any ring-like distribution of DiI localizations (Figs. 1 and 2).

When the membrane is tiled 0° or 90° relative to the coverslip, the anisotropic emission of DiI does not contribute to a systematic shift of the localizations. This is particularly relevant to PLM images of membrane tubules in which most of the membrane is parallel or perpendicular to the coverslip. Accordingly, the pPLM images of membrane tubules displayed a clear ring-link distribution (Fig. 4). The expected localizations for CTxB bound to the outer leaflet of a tubule should exhibit no detected localizations in the center of the tubule since CTxB cannot penetrate the membrane. Scare localizations of CTxB in the central part of the tubule was presumably due to the low stiffness and undulating motion of the tubule that enabled CTxB bound to the tubule exterior to be observed within the tubule center upon *z*-projection (Figs. 4 and 5).

**Membrane tension affects bud formation**

SLBs were made by rupturing GUVs upon a glass coverslip to create SLB patches with laterally varying tension. Near the center of an SLB patch, there was initially a local increase in membrane from a portion of the GUV or nested GUVs that did not become fused to the glass coverslip. These areas had a high concentration of DiI per sample volume but were removed with vigorous washing that yielded an apparently uniform SLB. However, membrane buds were most likely to form in the center of the SLB patch, close where the unfused vesicles were, rather than close to the edge of the SLB and the exposed glass coverslip (Fig. S3). The bud-forming region of the SLB typically extended >10 μm away from the center of the SLB batch, and the perimeter of the patch without bud formation was commonly 5 ± 4 μm wide. We interpret this as a variation in membrane tension and availability of extra membrane for topological changes.

Variations in SLB tension could occur during two distinct stages of SLB creation during GUV fusion. The first stage would consist of the initial GUV-glass contact, before or immediately after the GUV has ruptured. The rupture of the GUV may have exposed a loose or floppy bilayer to the glass coverslip and initiated a membrane-glass contact that trapped nanoscale undulations and decreased lateral tension in the SLB. Over time, the bilayer would spread across the glass with Marangoni flow, as encouraged by the membrane-glass adhesion, and yielded higher lateral membrane tension, similar to as seen previously (34). Accordingly, the center of each SLB patch would have a lower membrane tension and encourage bud and tubule formation, as observed.

The buds formed at random locations, there existed no nearest-neighbors persistent distances between the buds. Buds initially formed at the central part of the ruptured GUV of low membrane tension. Keeping a constant membrane bud density within a given area, more buds formed with time at random locations propagating radially outward into relatively higher tension regions.

**Membrane budding slows CTxB and DiI diffusion**

The diffusion of CTxB on membranes has been reported with widely varying rates, including rates that range from 0.04 to 2.44 $\mu m^2$/s within a single cell, and spatially confined in regions that are 100 to 1800 nm diameter (35, 36). Even in the absence of coexisting lipid phases such as > 99% DOPC model membranes, CTxB exhibited multiple diffusion populations, with one population of $D = (0.18 \pm 0.04)$ $\mu m^2$/s and the second population of $D = (0.06 \pm 0.02)$ $\mu m^2$/s with transient confinement in regions as small as 20 nm radii (16, 17). These prior measurements had no means of detecting changes to membrane topography or correlating topography and mobility, which is a focus of this manuscript. The diffusion measurements reported here are consistent with both the previously reported diffusion rates and confinement sizes while demonstrating local membrane curvature as a mechanism yielding variation in CTxB behavior in a membrane. This curvature-dependent analysis of CTxB diffusion and accumulation has the potential to explain prior measurements of both distinct populations of CTxB diffusion rates (Fig. 6) and the inter-membrane molecular sorting that is independent of lipid phase. Variation in CTxB diffusion rates and curvature-based CTxB sorting may reflect a variation in the number of bound GM1 to the CTxB molecule.

The mechanisms by which membrane bending slows CTxB and DiI diffusion are most likely to be the result of molecular crowding, local phase separation, and/or a curvature-dependent membrane viscosity. If CTxB becomes dense enough in a local region of the bilayer, it would be expected that this crowding would slow the diffusion of CTxB or lipids (*i.e.*, DiI) within the membrane. Additionally, it is feasible that the local concentration of GM1 was increased sufficiently as to drive the local lipid environment into a more ordered state and cause an increase in the effective membrane viscosity, as would be expected for more ordered lipid environment (37–39). It is not likely that the 99.4 mol% POPC bilayer would have significant phase separation, but CTxB-encouraged GM1 accumulations are possible.

Finally, it is feasible that the membrane bending itself affects the local effective lipid viscosity and the free diffusion of lipids and proteins through the membrane buds. This effect was observed for SLBs draped over nanoparticles to create controlled membrane bending-dependent on the size of the nanoparticle (40). Effective membrane viscosity changes due to the membrane bending. This mechanism would influence the diffusion of the DiI molecules in the bottom leaflet that is against the substrate or on the inside of the bud, on which there is no bound CTxB. Prior reports have demonstrated how membrane bending is able to slow lipid diffusion without the incorporation of GM1 or CTxB (40) [cite PLM].

**Bud formation and molecular sorting does not require lipid phase separation**

SLBs of >99% POPC or DPhyPC were used for these studies due to their strongly disordered acyl tails and the minimal possibility of phase separation with 0.3% GM1 and 0.3% DiI included. The liquid-to-gel transition temperatures for POPC and DPhyPC are -2°C and <-120°C, respectively. To further minimize the possibility of a spontaneous GM1 clustering in the bilayers, a low GM1 concentration was used (41). Prior studies that reported sorting of CTxB to the neck of membrane buds required a ternary mixture of cholesterol, sphingomyelin, and POPC to observe CTxB sorting (22). However, curvature in these prior studies was of a significantly larger radius of curvature (μm scale) such that the curvature-dependent sorting was presumably weaker than that observed here (nm scale).

Even though lipid phase separation is not likely in these experiments, it is feasible that nanodomains of GM1 were formed and were stabilized by the multivalent CTxB binding. This possibility is considered in later discussions of local molecular mobility and the mechanisms of CTxB-induced membrane bending.

**Bud formation is energetically feasible**

The spontaneous bud nucleation and tubulation are controlled by the membrane bending rigidity, density of CTxB, and the adhesion of the SLB with the substrate. For bud formation to be spontaneous, the energy released by CTxB-GM1 binding must be greater than the energy put into bending the membrane and separating the membrane from the glass. The total energetic cost of bending the membrane ($E_{Bend}$) was estimated via Helfrich energy model (42) for at 50 nm top radius of curvature hemispherical bud with a 20 nm radius of curvature collar smoothly connecting the bud to the surrounding planar SLB (Fig. 1B), such that

$$E_{Bend} = \int (\kappa (H - H_0)^2 + \bar{\kappa} K) dA \ . \tag{Eq. 3}$$

This incorporates the membrane bending rigidity ($\kappa$), membrane Gaussian curvature modulus ($\bar{\kappa}$), the mean local membrane curvature ($H$), the local Gaussian curvature ($K$), the intrinsic membrane curvature ($H_0 \approx 0$), and the area of the bud ($A$). The bending rigidity of a POPC membrane in TRIS buffer at T = 22 C is $\kappa = (12.9 \pm 0.4)$ x $10^{-20}$ J and $\bar{\kappa} \approx \kappa$ (43, 44). The energy required to bend the membrane into the presumed configuration was calculated analytically $E_{Bend}$ = $(5 \pm 1)$ x$10^{-18}$ J.

The adhesion energy of the bilayer to the glass substrate is given by $w = 10^{-8}$ J/m$^2$ (45). A bottom radius of 67 nm yields 1.4 x $10^4$ nm$^2$ of the SLB to be separated from the substrate, the energy cost of lifting the membrane off the substrate ($E_{Adhesion}$) equals $(1.4 \pm 1)$x $10^{-22}$ J, which happens to be smaller than $k_B T$.

The intrinsic free energy released per CTxB binding to the GM1 in the SLB is equal to -67 $\pm$ 2 kJ/mol (46). The area of the CTxB pentamer is equal to 106 nm$^2$ and there would be space for 120 CTxB to bind to just the neck region of this membrane bud (10, 23). Accordingly, the energy change upon CTxB accumulating around the neck of this nanoscale bud ($E_{Bind}$) is (-1.3 $\pm$ 0.5) x$10^{-17}$ J.

In comparing these three energetic components, we found | $E_{Bind}$ | > | $E_{Bend}$ + $E_{Adhesion}$ |. In conclusion, there is ample energy from CTxB binding to drive nanoscale bud formation spontaneously. However, further discussion is warranted to consider how much of this energy of CTxB binding is expected to drive membrane bending, and if CTxB binding to the bud neck with a negative Gaussian curvature is consistent with prior studies.

**The forces that drive budding**

A complex interplay of factors could contribute to the budding process such as 1) CTxB steric crowding; 2) CTxB insertion into the membrane (47); 3) CTxB-GM1 cross-linking; 4) the positive intrinsic curvature of GM1 (48); 5) GM1 clustering and lipid phase separation (41); 6) the extended long acyl chain of GM1 causing wedging in the opposing bilayer leaflet (16), 7) and the asymmetric GM1 concentrations in the bilayer leaflets. The steric pressure between the crowded CTxB within a nanoscale area could inherently encourage membrane bending if there was an attractive force between the GM1 to counter the steric repulsion between CTxB and provide a local membrane torque (49). An attraction between GM1 is plausible considering the strong liquid-ordered phase preference of GM1 and the possibility of GM1 unbound to CTxB further accumulating around the clusters of CTxB-GM1. Both a GM1-GM1 clustering and CTxB binding preferentially on one of the bilayer would encourage the accumulation of GM1 on the top leaflet. This leaflet mismatch of GM1 concentration could encourage bending due to the intrinsic molecular shape of GM1. Further, the molecular shape of CTxB itself is likely to encourage a negative membrane curvature. The molecular shape of CTxB and Shiga toxin both have glycolipid binding pockets that are elevated above from the bottom of the folded protein, the toxin-lipid binding encourages a penetration of the protein into the bilayer and/or a local wrapping of the membrane around the protein, as has been most explicitly shown for Shiga toxin (47).

Finally, the multivalent binding of CTxB may be critical for the preference of CTxB to bind to negatively curved membranes, such as regions of negative Gaussian curvature over planar membranes. Since CTxB is not typically saturated by binding to five GM1 simultaneously, the rotational asymmetry of the occupied GM1 binding pockets on the CTxB may result in a shifted CTxB preference to bind to one dimension of negative curvature and have a minimal preference for the membrane curvature in the other principal curvature dimension. Accordingly, if presented with a negatively curved membrane, the CTxB bound to ~3 GM1 would be stable at any angle. However, if presented with negative Gaussian curvature, the CTxB bound to ~3 GM1 would require it to be rotated such the GM1 are spaced along the dimension of negative curvature, limiting the degrees of freedom for CTxB orientation on the membrane but still occupying the GM1 binding pockets and being appropriately wrapped by the bilayer. Further experimentation to clarify the mechanisms of CTxB-induced membrane curvature is currently underway.

**Membrane curvature induced CTxB sorting**

Studies regarding CTxB location at certain membrane curvature gradients have shown that CTxB preferentially localizes at negatively curved membrane regions (3) and at the neck of micron-scale engineered membrane buds (22). Here, CTxB were observed to initially cluster and spontaneously form small membrane buds (Fig. 1A, *black arrows*). As the buds grew in size, the accumulation of CTxB around the perimeter of larger membrane protrusions became increasingly apparent (Fig. 1A, *white arrow*). CTxB also localized at similar membrane geometry on bilayer draped over a 70 nm radius nanoparticle (Figs. 7 and S2).     Thus,     the 'ring-like' shape of localizations obtained for CTxB demonstrates the preferential location of CTxB at the neck of the curvature. This observation falls in agreement with previous reports that observed the accumulation of CTxB at the neck of lithographically patterned spherical membrane protrusions with 25 μm diameter that was dependent on coexisting liquid lipid phases (22). By observing 20x smaller curvatures here, the curvature-based sorting forces were presumably larger and not requiring assistance by lipid phase separation for CTxB sorting to occur.

**CONCLUSION**

PLM-enabled the direct observation of CTxB and membrane dynamics along with the budding events driven by CTxB. In this manuscript, PLM was used to reveal multicolor super-resolution images of CTxB and the induced bud growth on a supported, quasi-one component lipid bilayers. This demonstrated CTxB ability to induce and form nanoscale membrane curvature. Our data provide context to prior studies with CTxB that observed time-dependent diffusion rates and diverse internalization mechanisms. We demonstrated that the molecular mobility of CTxB and DiI are affected by the nanoscale membrane structures induced by CTxB. On planar SLBs, the diffusion rates of DiI and CTxB of $(0.39 \pm 0.05)$ μm$^2$/s and $(0.09 \pm 0.02)$ μm$^2$/s, respectively, fall in agreement with previous reports on cells and synthetic membranes. However, the diffusion coefficients at the center of the induced buds are $(86 \pm 10)\%$ and $(76 \pm 4)\%$ less than that on the planar membranes for DiI and CTxB, respectively. DiI and CTxB underwent transient confinement in regions that later appeared to be nanoscale protrusions as small as 30 nm radius. Our studies demonstrated the budding process was reversible and dependent on CTxB concentration. PLM will aid in providing new information for previously untestable nanoscale processes coupled with changes in membrane topography. We propose a foundational mechanism of CTxB trafficking in cells dependent on the spontaneous membrane curvature induced by CTxB. In further manuscripts, we will explore the effect of changing the GM1 structure and membrane composition on the budding process, as well as using mutant, monovalent CTxB that binds to one GM1 (50). Biological functions of the cell are dictated by the sorting, mobility, and organization of its constituents that affect the structure of the cell membrane facilitate diverse essential membrane processes.

**ACKNOWLEDGEMENTS**

The authors thank Xinxin Woodward, Dipanwita De, Rebecca Meerschaert, and Eric Stimpson for valuable discussions. A.M.K. was funded by Thomas C. Rumble Fellowship Award. Financial support was provided by Wayne State University laboratory startup funds, Richard J. Barber, and a CAREER award from the National Science Foundation (DMR1652316).

**AUTHOR CONTRIBUTIONS**

A.M.K. and C.V.K. designed the experiments, analyzed the data, and prepared the manuscript. A.M.K. performed the experiments.

**COMPETING FINANCIAL INTERESTS STATEMENT**

The authors have no competing financial interests.

**FIGURES**

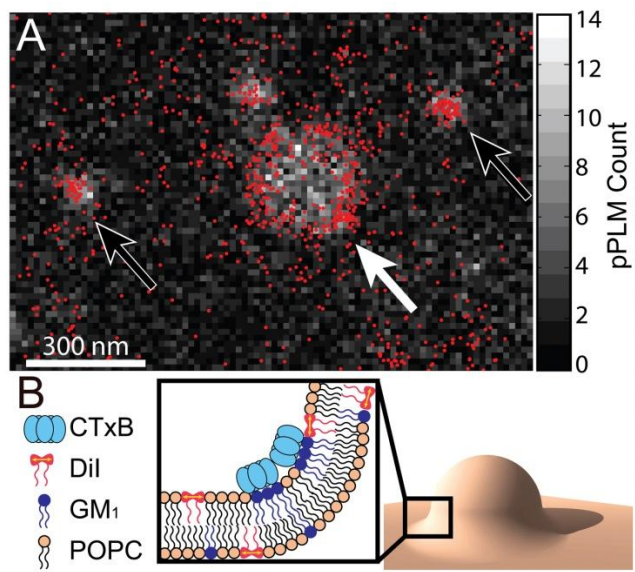

**FIGURE 1** Colocalization of CTxB accumulations and membrane bending on supported lipid bilayers demonstrate the inherent capability of CTxB to induce membrane curvature. (A) The 2D histogram of DiI localizations with p-polarized excitation highlights where the membrane is more perpendicular to the coverslip. This is overlaid with the single-molecule localizations of CTxB that are displayed with red points. The membrane is bending where CTxB is concentrated. Larger membrane buds, as indicated by a *white arrow,* shows the high concentration of CTxB preferentially at the neck of the bud, where the membrane has a negative Gaussian curvature, as shown schematically in (B). Smaller membrane buds, as indicated by *black arrows,* display a local accumulation of CTxB, although identification of where on the small buds CTxB is most concentrated is not feasible at this resolution.

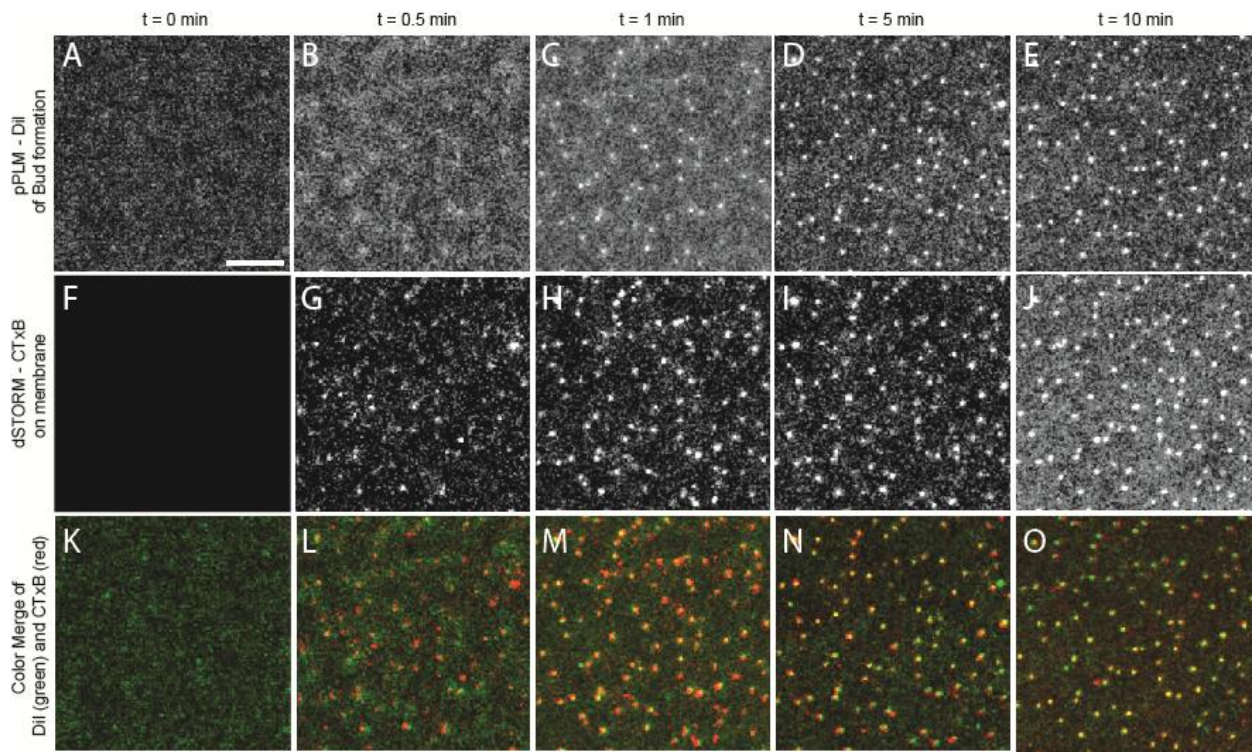

**FIGURE 2** Simultaneous observation of (A-E) membrane bending detected via pPLM and (F-J) CTxB clustering on the SLB detected via dSTORM reveals the CTxB-induced membrane bending. Before CTxB is added, (A) the DiI localizations by pPLM are uniform and (F) no CTxB localizations were found. Within the first 30 sec of CTxB on the SLB, (B) slight variations in the DiI localizations were present and (G) CTxB clusters formed. At later times, (C-E) the membrane buds become increasingly apparent and (H-J) CTxB became increasingly concentrated at the buds.(K-O) Color merge for the membrane (A-E) in green with CTxB (F-J) in red. Scale bar represents 200 nm.

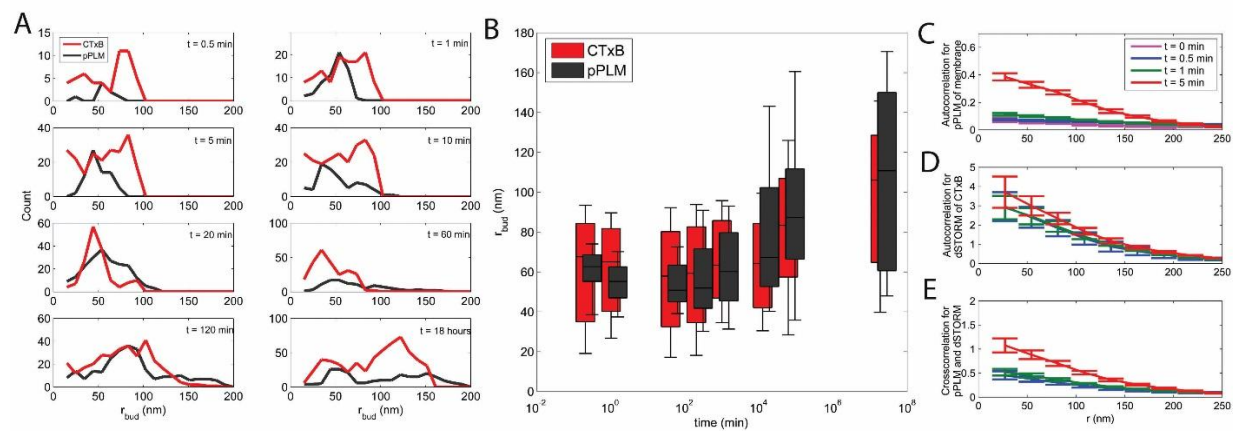

**FIGURE 3** The super-resolution capabilities of PLM and dSTORM reveal membrane bud size ($r_{bud}$) versus CTxB incubation time. As some buds grow to a larger diameter or into tubules (Fig. 4), new small buds ($r_{bud} < 50$ nm) continue to form and the distribution of $r_{bud}$ widens over time. (A) The number and size of the induced buds increase as the CTxB incubation time with the membrane increases. (B) Whisker plot for the sizes of the nanoscale buds detected in PLM and dSTORM for the membrane and CTxB, respectively. (C,D) Autocorrelation analysis of DiI and CTxB as a function of incubation time, respectively. (E) Cross-correlation analysis of DiI and CTxB as a function of time.

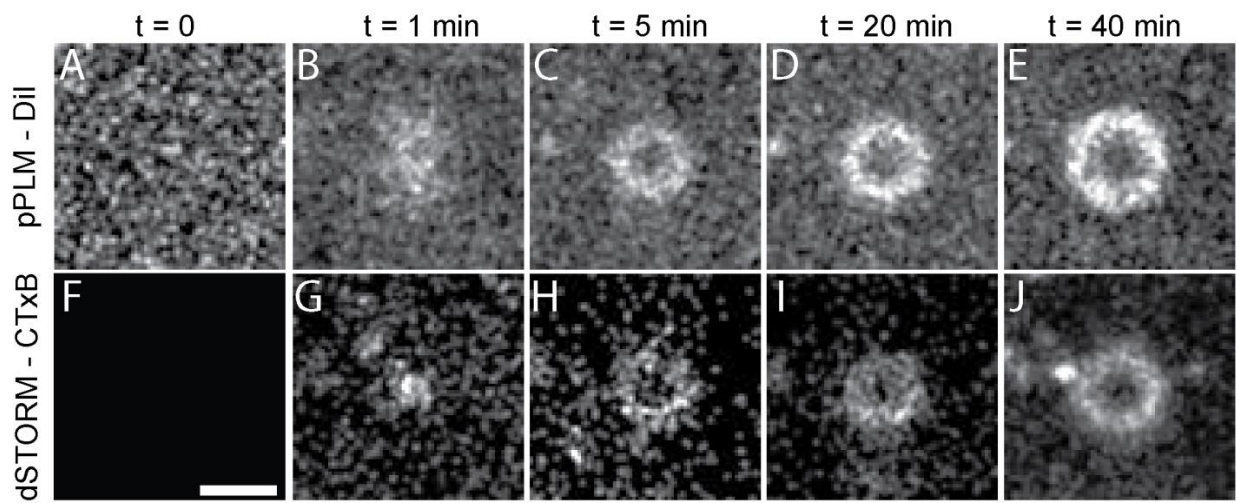

**FIGURE 4** Some membrane buds grow into tubules extending away from the glass coverslip (Fig. S4). These features start with (B) a membrane bud and (G) small clusters of CTxB ($r_{bud} <$ 100 nm). Over time, (C) a ring of DiI localizations forms as the bud top extends farther from the coverslip and (D, E) the ring widens with an increasing tubule diameter. This membrane bending is driven by (H-J) the colocalization of CTxB at the base of the tubule where the negative Gaussian membrane curvature is present.

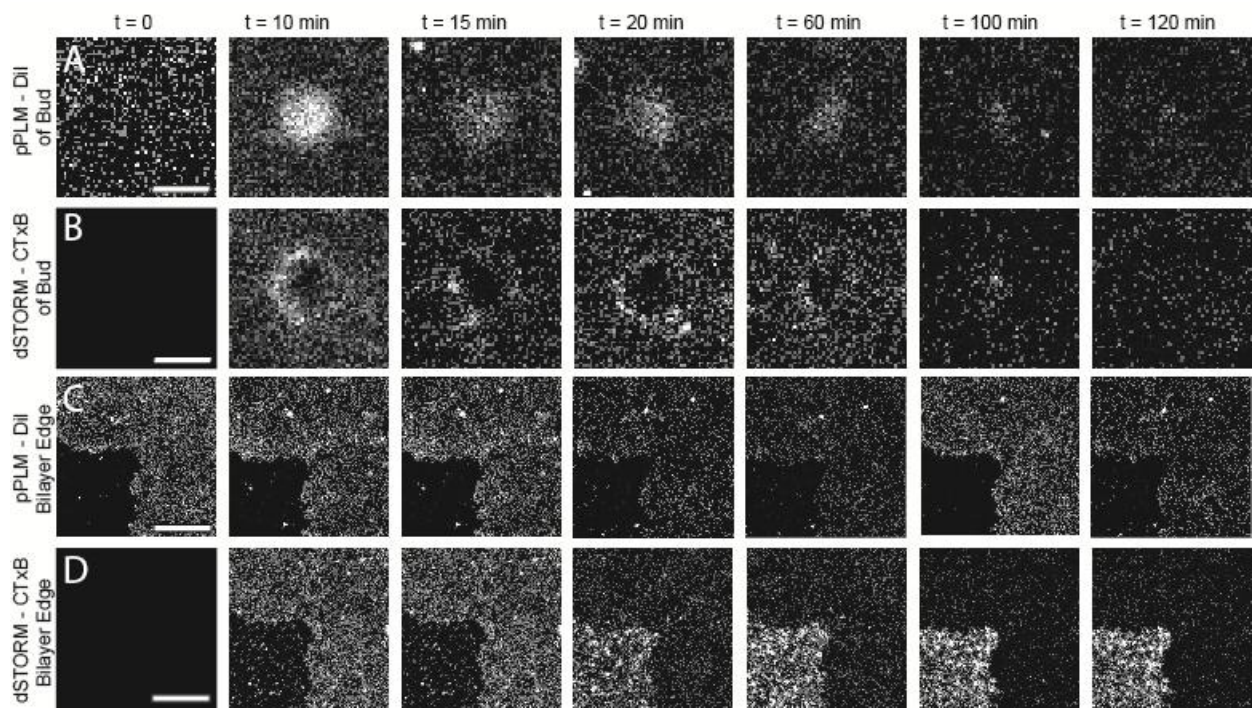

**FIGURE 5** CTxB depletion from the membrane showed that the budding process is dependent on CTxB and reversible. In this experiment, the surrounding glass coverslip was prepared as to encourage CTxB absorption and removal from the SLB. (A) The pPLM images show the membrane buds decreasing in height and diameter.  (B) The dSTORM images of CTxB show a decrease of CTxB on the SLB and a uniform concentration across where the bud had been. (C) represents the time-lapse of the membrane edge surrounded by the treated glass. With time, the bilayer showed no change except for bud disappearance. (D) However, [CTxB] on the bilayer dramatically decreases at t = 15 min, while [CTxB] on glass increases. This is due to the sticking of the diffusing CTxB to the glass after close proximity. Scale bars in (A,B) represents 100 nm and in (C,D) 1 μm

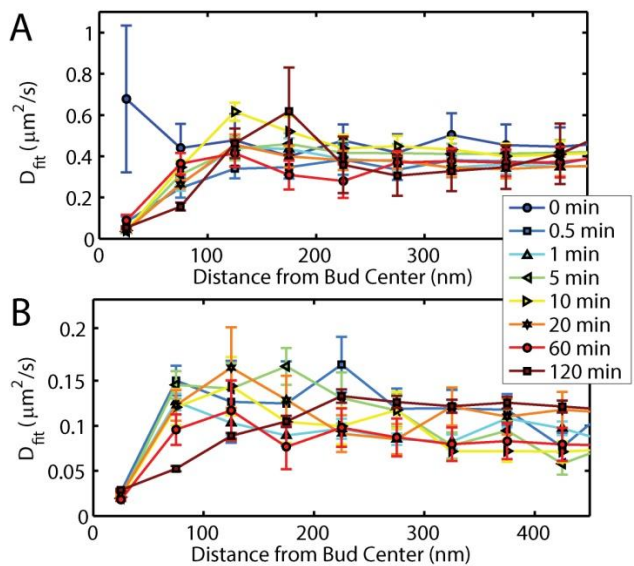

**FIGURE 6** Single-particle tracking was performed on both (A) DiI and (B) CTxB as a function of position within a membrane bud. Before the CTxB is added ($t = 0$ min), no CTxB was located on the SLB, no membrane buds were present, and random locations were chosen for the pPLM analysis to confirm our analysis routines demonstrated no significant variation in $D_{fit}$ versus distance away from the randomly chosen locations. At all later times, a significant slowing of both the DiI and CTxB diffusion is observed within 50 nm of the bud center. The error bars represent at a 95% confidence interval of the fitting of Eq. 1 to the histograms of step lengths that were binned based on the distance from the bud center of the mean of the two linked localizations.

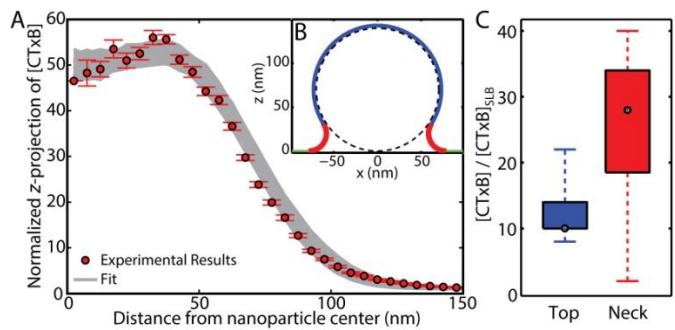

**FIGURE** 7 When membrane buds were formed by draping SLBs over nanoparticles of known size (70 nm radius), the sorting of CTxB versus membrane curvature could be determined. (A) Many combinations of the fitting parameters yielded quality fitting to the experimental data. (B) The membrane topography over the nanoparticle could be approximated to connect the concentration of CTxB per membrane area to the acquired $z$-projected data. The distribution of adequate model fits to the experimental data yielded a mean and standard deviation of CTxB concentration on the membrane top and neck relative to the planar SLB was (12 ± 4) and (26 ± 11), respectively, with the median, quartiles, and range of CTxB concentrations are shown in (C).